\documentclass[usegraphicx,usenatbib]{mn2e}

\usepackage{natbib}
\bibpunct{(}{)}{;}{a}{}{,}
\usepackage{aasmacros}

\newcommand{\kms}{\mbox{\,km\,s$^{-1}$}}

\newcommand{\eg}{e.g.\ }

\newcommand{\Msun}{M_{\odot}}

\newcommand{\vph}{$v_{\rm ph}$}

\newcommand{\OI}{O~{\sc i}}

\newcommand{\CI}{C~{\sc i}}

\newcommand{\NaI}{Na~{\sc i}}
\newcommand{\MgII}{Mg~{\sc ii}}
\newcommand{\MgI}{Mg~{\sc i}}

\newcommand{\SiII}{Si~{\sc ii}}
\newcommand{\SiIII}{Si~{\sc iii}}
\newcommand{\SII}{S~{\sc ii}}
\newcommand{\CaII}{Ca~{\sc ii}}
\newcommand{\TiII}{Ti~{\sc ii}}
\newcommand{\TiIII}{Ti~{\sc iii}}
\newcommand{\CrII}{Cr~{\sc ii}}
\newcommand{\CrIII}{Cr~{\sc ii}}

\newcommand{\FeII}{Fe~{\sc ii}}
\newcommand{\FeIII}{Fe~{\sc iii}}
\newcommand{\CoII}{Co~{\sc ii}}

\newcommand{\NiII}{Ni~{\sc ii}}
\newcommand{\Feff}{$^{54}$Fe}
\newcommand{\Fefs}{$^{56}$Fe}
\newcommand{\Cofs}{$^{56}$Co}
\newcommand{\Nifs}{$^{56}$Ni}
\newcommand{\Nife}{$^{58}$Ni}

\newcommand{\DeltaB}{$\Delta m_{15}(B)$}

\def\lsim{\mathrel{\rlap{\lower 4pt \hbox{\hskip 1pt $\sim$}}\raise 1pt\hbox {$<$}}}
\def\gsim{\mathrel{\rlap{\lower 4pt \hbox{\hskip 1pt $\sim$}}\raise 1pt\hbox {$>$}}}

\title[Abundance stratification in Type Ia Supernovae - \\
	II: SN\,2004eo]{Abundance stratification in Type Ia Supernovae - \\
	II: The rapidly declining, spectroscopically normal SN\,2004eo}

\author[Mazzali et al.]
{Paolo A.~Mazzali$^{1,2,3,4}$\thanks{E-mail: mazzali@mpa-garching.mpg.de},
D.N. Sauer$^1$, A. Pastorello$^5$, S. Benetti$^6$, W. Hillebrandt$^1$ \\
$^1${\it Max-Planck Institut f\"ur Astrophysik,
  Karl-Schwarzschild-Str. 1, 85748 Garching, Germany}\\
$^2${\it Istituto Naz. di Astrofisica-Oss.\ Astron., Via Tiepolo, 11,
  34131 Triste, Italy}\\
$^3${\it Research Centre for the Early Universe and Dept. of Astronomy,
School of Science, Univ. of Tokyo, Bunkyo-ku, Tokyo 113-0033, Japan}\\
$^4${\it Kavli Institute for Theoretical Physics, University of California,
Santa Barbara, CA 93106-4030, USA} \\
$^5${\it Astrophysics Research Centre, Queen's University, Belfast BT7 1NN,
UK} \\
$^6${\it Istituto Naz. di Astrofisica-Oss.\ Astron., vicolo dell'Osservatorio,
5, 35122 Padova, Italy }
}
\date{Accepted 2008 March 5. Received 2008 March 3;
in original form 2008 January 30}

\volume{000}

\pagerange{1\,-\,12} \pubyear{2008}

\begin{document}
\maketitle

\label{firstpage}

\begin{abstract}

The variation of properties of Type Ia supernovae, the thermonuclear explosions
of Chandrasekhar-mass carbon-oxygen white dwarfs, is caused by different
nucleosynthetic outcomes of these explosions, which can be traced from the
distribution of abundances in the ejecta. The composition stratification of the
spectroscopically normal but rapidly declining SN\,2004eo is studied performing
spectrum synthesis of a time-series of spectra obtained before and after
maximum, and of one nebular spectrum obtained about eight months later.
Early-time spectra indicate that the outer ejecta are dominated by oxygen and
silicon, and contain other intermediate-mass elements (IME), implying that the
outer part of the star was subject only to partial burning. In the inner part,
nuclear statistical equilibrium (NSE) material dominates, but the production of
\Nifs\ was limited to $\sim 0.43 \pm 0.05 \Msun$. An innermost zone containing
$\sim 0.25 \Msun$ of stable Fe-group material is also present.  The relatively
small amount of NSE material synthesised by SN\,2004eo explains both the
dimness and the rapidly evolving light curve of this SN.

\end{abstract}



\section{Introduction}

Type Ia supernovae (SNe\,Ia), the thermonuclear explosions of carbon-oxygen
white dwarfs that approach the Chandrasekhar mass limit \citep[for a review,
see][]{HillNie00},  are characterised by a simple optical light curve, showing
a rise to maximum followed by a decline. The light curve is governed by the
deposition of the $\gamma$-rays and the positrons produced by the decay of
\Nifs, which is copiously produced in the explosion and decays into \Cofs\ and
hence \Fefs\ \citep{arn82,kuch94}, and by the diffusion in the expanding ejecta
of the optical photons that are produced in the thermalisation process.

A correlation between the peak luminosity of SNe\,Ia and the shape of the light
curve \citep{phil93,riesspresskirsh96,phil99}, makes this class of SNe good
standardizable candles. SNe\,Ia have been used to explore the expansion
properties of the Universe out to about one third of its present age
\citep{riess98,perl99}.

While the SN luminosity is related fairly directly to the amount of \Nifs\
ejected \citep{arn82}, so that the observed spread in peak luminosity suggests
a corresponding spread in \Nifs\ mass of about one order of magnitude
($\sim 0.1$ to $\sim 1 \Msun$) \citep{cont00,stritz06,maz07a}, the light curve
width depends primarily on the optical opacity in the ejecta, which is a
function of both temperature and composition \citep{hmk93}. In the H-free
ejecta of SNe\,Ia, line opacity dominates over continuum processes
\citep{karp77,hmk93,paul96}. Heavier elements have a thicker array of lines
which, in combination with the rapid and differential expansion of the SN
ejecta, causes the effective frequencies of the lines to overlap over the
volume of the ejecta, efficiently blocking most UV and blue photons. For most
of these photons the only way to escape is to be re-emitted in some less
saturated transition in the red \citep{p&e00,m00}. This ``fluorescence''
process determines the shape of Type I SN spectra at early times, when the
density is sufficiently large. \citet{maz01} showed that if the opacity is
parametrised according to the average number of line transitions of the species
that dominate the ejecta (singly and doubly ionised Si, S, Fe, Co and Ni for
SNe\,Ia), SNe\,Ia that produce more \Nifs\ are not only brighter but also have
a larger opacity and hence broader light curves.

Additionally, different conditions within the white dwarf (\eg metallicity) can
lead to different nucleosynthesis of Fe-group isotopes \citep{timbt03}. For the
same production of NSE material, SNe with a higher fraction of stable isotopes
(\eg \Feff, \Nife) over radioactive \Nifs\ have light curves with a similar
shape but different brightness. This may cause a spread in the observed
luminosity-light curve shape relation \citep{m&p06}.

Although an explanation for the observed behaviour of SNe\,Ia seems to have
been found, a conclusive proof of the nature of the physical mechanism behind
these explosions still eludes us. Subsonic deflagrations of Chandrasekhar-mass
carbon-oxygen white dwarfs seem to be unable to produce the required energy and
\Nifs\ yield \citep{roepke07}. Detonations incinerate the white dwarf to NSE,
and do not produce the IME that are observed in the spectra. A transition from
a deflagration to a detonation \citep{khok91} is a possible solution
\citep{maz07a}, but its physical basis remains to be established, despite much
effort \citep{roepniem07,roep07,woos07}.

An alternative approach to the study of SNe\,Ia, rather than computing
explosions and predicting their appearance, is to map the outcome of real
events. This can be done performing spectrum synthesis of a time-series of SN
spectra, following the progressive exposure of deeper and deeper layers of the
ejecta as they expand.

Spectra at early times, in the so-called photospheric phase, probe only the
outer half or so of the depth of the ejecta. A super-nebular phase then
follows, during which the position of the pseudo-photosphere is extremely
wavelength dependent and some features already follow nebular physics. A more
direct and complete view of the inner ejecta can be obtained about one year
after the explosion, when the gas in the young SN remnant is heated by
collisions with the fast particles from the deposition of  $\gamma$-rays and
positrons and cools by emission of radiation, mostly in forbidden [\FeII] and
[\FeIII] lines. In this phase, the dominance of Fe lines testifies that \Nifs\
is synthesised mostly in the deepest parts of the star, in line with
theoretical expectations \citep[\eg][]{iwa99}. A relation between the width of
the Fe emission lines - which indicates the velocity of expansion of the bulk
of \Nifs\ - and both the SN peak brightness and the light curve decline rate
\citep{maz98} confirms that in more luminous SNe a larger fraction of the star
is burned to NSE. It is therefore essential that SNe that have been monitored
in the peak phase are also observed spectroscopically and photometrically in
the nebular phase.

Only few examples of SNe\,Ia with very detailed time coverage were available
\citep[\eg][]{salvo01} before the European Research and Training Network on the
physics of Type Ia Supernovae (RTN) began collecting data of nearby SNe\,Ia
systematically \citep[\eg][]{ben04}.  Now, the availability of a sizable sample
of SNe\,Ia with detailed observational coverage makes it possible to study in
greater depth the properties of SNe of different luminosities. Since such
differences are the result of different nucleosynthetic outcomes of the
explosions, deriving the composition of the ejecta from the data should shed
light on the details of the explosion process and it variations.

After initial efforts using homogeneous compositions \citep[\eg][]{maz93}, the
more computationally expensive analysis of a time series of spectra using
stratified abundances has been attempted in only one case so far.
\citet[][hereafter Paper I]{steh05} mapped the ejecta of the normally luminous
SN\,2002bo from the centre out to $v \sim 20,000$\,\kms. They showed that
abundances are stratified but some mixing does occur. They detected burning
products, namely stable Fe, Si and Ca, at very high velocities. This material
gives rise to high velocity features (HVF), an apparently ubiquitous property
of SNe\,Ia at the earliest phases \citep{maz05}. For SN\,2002bo, the IME region
extends down to $\sim 10,000$\,\kms, below which NSE material dominates.

This distribution of elements in SN\,2002bo seemed to favour a Delayed
Detonation scenario. When the abundances derived in this sort of tomographic
scan of the ejecta were used to generate opacities in a $\sim 10^{51}$\,erg,
Chandrasekhar-mass explosion with the density structure of the classic W7 model
\citep{nom84}, the observed bolometric light curve was reproduced extremely
well, including the early rise. This is induced by mixing out of \Nifs, which
is not predicted by most current explosion models but is indeed detected in
early SN\,Ia spectra \citep{tan08}.

In this series of papers our aim is to describe the composition of the ejecta
of SNe\,Ia with different luminosities and light curve decline rates.
SN\,2002bo was a SN\,Ia of average luminosity, which synthesised
$\sim 0.5 \Msun$ of \Nifs.  In this second paper of the series, we address a SN
at the dim end of the distribution of spectroscopically normal SNe\,Ia.
SN\,2004eo was observed by the RTN \citep{pasto07}. It was a relatively
underluminous event, with a $B$-band post-maximum decline rate \DeltaB$= 1.46$\,mag,
which makes it very similar to the template SN\,1992A. Like many such events,
SN\,2004eo showed a slow evolution of the \SiII\ line velocity \citep{ben05},
measurements of which indicate that Si extended down to layers at $\sim 8500$\,\kms,
which is deeper than in SN\,2002bo.  The earliest spectra show a mild  HVF in
the \CaII\ IR triplet, and possibly in \SiII\ 6355\,\AA.

The spectra and the light curve of SN\,2004eo were presented by
\citet{pasto07}.  The sample, although temporally not as dense as for
SN\,2002bo, covers both the peak and the nebular phases, and is probably the
best so far for a SN\,Ia of this luminosity group, offering a complete view of
the ejecta of SN\,2004eo except perhaps for the highest velocities.

In Section 2 our modelling is presented, and its limitations assessed. In
Section 3 the modelling of the photospheric-epoch spectra is described, while
the modelling of the nebular spectrum is discussed in Section 4. In Section 5
the abundance stratification obtained from the models is presented and
discussed. In Section 6 we present and discuss a theoretical light curve
computed on the basis of the abundances that have been derived. Finally, in
Section 7 we give our conclusions.

\section{Method}

Our modelling strategy of is to follow the evolution of the SN in the early
phases with a photosphere-plus-expanding-envelope code based on the Sobolev
approximation, and to explore the innermost part of the ejecta modelling a
single nebular spectrum taken $\sim 1$ yr later with a non-local thermodynamic
equilibrium (non-LTE) nebular code.

For the early-time spectra, we used a Montecarlo code based on that first
described by \citet{m&l93} and updated by \citet{l99} and \citet{m00}. The code
assumes a sharp photosphere, where the luminosity is emitted, and adopts the
Sobolev approximation to follow energy packets in their propagation through the
SN ejecta, which are represented by a density-velocity structure. The emerging
luminosity $L$, the velocity of the pseudo-photosphere \vph, and an epoch $t$
since the explosion are required input, as is a set of depth-dependent
abundances. Energy packets representing photon batches can interact with the
ejecta via either line absorption or electron scattering. Upon absorption, a
packet can be reprocessed into a different line, thus describing the process of
line blocking which is essential for spectrum formation in Type I SNe
\citep{m00}. No continuum processes are assumed to occur above the
pseudo-photosphere. The properties of the radiation field and of the gas are
iterated to convergence, temperature structure is established in radiative
equilibrium and the ionization and excitation structures are computed using a
modified nebular approximation that describes deviations from local
thermodynamic equilibrium \citep[for details, see][]{m&l93}. We use the code
version that employs abundance stratification, as discussed in Paper I.

The nebular spectrum has been modelled using a code that follows in 1D the
propagation and the deposition of the $\gamma$-rays and positrons produced in
the decay of \Nifs\ into \Cofs\ and hence \Fefs\ using a Montecarlo scheme
\citep{capp97}. Gas heating by the fast particles created upon $\gamma$-ray and
positron deposition and the consequent cooling via line emission are described
in non-LTE \citep{axe80,rll92,maz01}.  Stratification in density and abundance
is used, as in \citet{maz07b}.

Before we describe our results in detail, it is necessary to discuss the
possible limitations of this method. Extracting the element distribution from a
series of SN spectra is an ill-posed inverse problem: too many of the
parameters that are used to compute the synthetic spectra are not known. In
particular, assuming a certain density distribution artificially narrows down
the number of possible solutions. Here we adopt the density-velocity
distribution of W7 \citep{nom84}, for both the early and late-time modelling.
Although W7 describes a more luminous SN\,Ia than SN\,2004eo, it represents a
typical density structure. Possible uncertainties related to using that
particular model are discussed in Section 7.  On the other hand, using any
other explosion model is no better method, as the uncertainties related to the
particular model would simply compound with those related to the spectrum
synthesis codes.

Additional caveats stem from the design of the code that we use to model the
photospheric-phase spectra. The use of a sharp photosphere is very effective,
and it keeps the analysis free from the details of the explosion model, but it
does not correspond to reality, in particular at times after maximum, when the
photosphere resides within the \Nifs\ zone and consequently a significant
fraction of the $\gamma$-ray deposition occurs above the photosphere. These
photons may alter both the flux level and the line profiles.  Fortunately, this
is mitigated by the decreasing density, which guarantees that even at advanced
stages most deposition still occurs below the photosphere.

Additionally, since the nebular spectra do not suffer from this problem, and
since the velocities sampled by the nebular lines exceed 7000\kms, combining
early photospheric-epoch results for the outer ejecta and late nebular ones for
the inner ejecta we effectively avoid being too affected by the lesser accuracy
of the post-maximum models.

Another shortcoming is the fact that we use a one-dimensional approach, while
in reality SNe\,Ia ejecta may be affected to some extent by aspherical effects.
However, three-dimensional spectral modelling can only be attempted starting
from an explosion model, or the number of free parameters would be impossibly
large, and it is a path that runs contrary to our approach.

When modelling spectra, it is practically impossible to establish objective
criteria with which to judge the goodness of a fit. In fact, it is not fair
simply to measure the deviation from the observed flux, because not all
wavelengths have the same weight when it comes to deciding whether the physical
quantitites that determine the spectrum are properly described. Deviations in
wavelength are if possible of greater importance, because they signal an
incorrect abundance, ionization or density distribution. Deviations in flux can
be due to an incorrect luminosity or temperature, but also to incorrect line
strengths, and these are again determined by density, ionization, and
abundances. Therefore, a human observer is probably the best judge of the
goodness of a fit.

In the case of a tomography experiment, it is not enough to fit a single
spectrum, but the entire series must be reproduced by a single set of input
data. There may be solutions that fit a single spectrum better than the one
finally adopted, but the overall result may not be as good as that obtained
with a different set. Therefore, a double compromise must be reached, on a set
of input parameters that give both acceptable individual fits and on a fair
reproduction of the spectral evolution.

Further details about the accuracy of our results are given in Sections 5 and 6.

\section{Photospheric-epoch spectra}

We have modelled 8 early-time spectra of SN\,2004eo, covering epochs from $-11$
to $+22$ days relative to $B$ maximum. There are later spectra
\citep[see][]{pasto07}, but we did not model them because the Montecarlo code
becomes progressively less accurate at advanced epochs and because they probe
very low velocities that are covered by the nebular spectrum.

We adopted a bolometric rise time of 19 days for SN\,2004eo. This is somewhat
less than the average $B$-band rise time \citep{riess99}, as motivated by the
fact that dimmer SNe\,Ia seem to rise somewhat more rapidly than brighter ones.
This is the consequence of the smaller opacity and the shorter photon diffusion
time in the less nuclearly processed ejecta of dimmer SNe \citep{maz01}. The
main input values for the synthetic spectra are recapped in Table 1, and the
synthetic spectra are shown and compared to the observed ones in Figures 1 and
2. Using this approach, we map the abundance distribution in the ejecta.  First
we briefly discuss each spectrum in turn.


\begin{table*}
\caption{Parameters of the synthetic spectra.}
\begin{tabular}{rrcrr}
\hline\hline
\noalign{\vspace{2pt}}
date~~~~ & epoch & $  \log L $      & $v_{\rm ph}$ & T$_B$ \\
     & (days)  &  [erg s$^{-1}$]  & (\kms)         & (K)   \\
\hline
19 Sept 2004 &  8.0 &  42.66& 12000 & 13791   \\
24 Sept 2004 & 12.7 &  43.03& 10400 & 13926  \\
27 Sept 2004 & 16.4 &  43.07&  9500 & 12424  \\
 2  Oct 2004 & 21.4 &  43.10&  9000 & 10621  \\
 7  Oct 2004 & 26.4 &  43.00&  7300 & 10060  \\
11  Oct 2004 & 30.4 &  42.85&  6000 & 9499   \\
14  Oct 2004 & 33.4 &  42.74&  4700 & 9746   \\
21  Oct 2004 & 40.4 &  42.53&  2800 & 10367  \\
\noalign{\vspace{2pt}}
\hline
\noalign{\vspace{2pt}}
\hline
\end{tabular}
\end{table*}


\subsection{19 September 2004: $-11$ days}

The earliest spectrum of SN\,2004eo has an epoch of $\sim 8$ days after
explosion (Figure 1a). Although the SN was not a very luminous one, at this
early epoch the small size of the ejecta results in a relatively high
temperature, as shown by the fact that the two Fe features in the blue, near
4300 and 4800\,\AA, are dominated by \FeIII\ lines. Other strong features are
due to \CaII, \SiII, \SII, \OI. The ratio of the two \SiII\ lines
($\lambda 5972$ and 6355\,\AA), which is a good temperature indicator
\citep{nug95} is rather large. As shown by \citet{hach08}, this is not directly
the effect of a low temperature, as then the ratio would show the opposite
trend, nor is it caused by \TiII\ lines, as claimed by \citet{garn04}, as in
that case the much stronger \TiII\ lines in the blue would completely change
the spectrum. It is the result of saturation of the 6355\,\AA\ line as \SiII\
becomes more abundant with respect to \SiIII\ going to lower luminosities,
hence lower temperatures.  So this is only indirectly the effect of
temperature. In the UV, the spectrum is blocked by \FeII, \FeIII, \NiII, \TiII,
\TiIII, \CrII, \CrIII\ and \MgII\ lines.  The flux redistribution caused by
line blocking determines the properties of the optical spectrum.


\begin{figure*}
  \includegraphics[width=144mm]{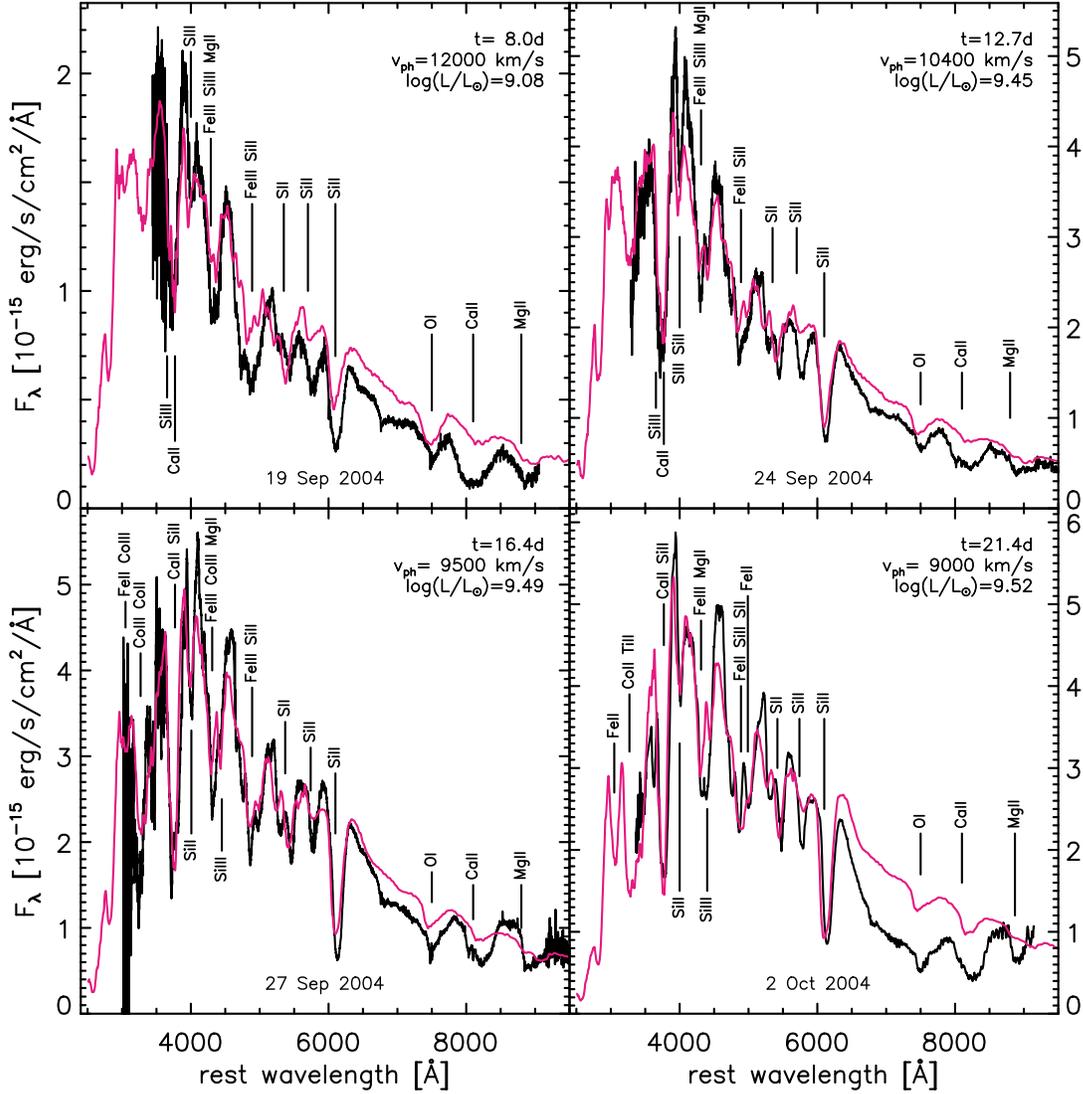} \caption{Fits to the earliest 4
  spectra of SN\,2004eo. Basic input parameters are shown and line
  identifications are marked. The spectra are as follows: {\bf a)} (top left):
  $t=-11$ days; {\bf b)} (top right): $t=-6$ days; {\bf c)} (bottom left):
  $t=-3$ days; {\bf d)} (bottom right): $t=+2$ days.}
\end{figure*}


\subsection{24 September 2004: $-6$ days}

The second spectrum has an epoch of 12.7 days after explosion (Figure 1b). The
increase in luminosity does not keep up with the increase in size, and the
temperature is slightly lower. Most lines and their strength ratios are however
similar to the previous epoch, since the temperature drop is small. The most
noticeable change is the increased strength of the \SII\ complex near
5500\,\AA\ ($\lambda \lambda 5433, 5454$\,\AA\ for the bluer feature,
$\lambda \lambda 5606, 5640$\,\AA\ for the redder one). This is indicative both
of recombination and of an increased sulphur abundance as deeper layers are
exposed. The abundance of silicon is also higher near the photosphere, but this
has less of an effect on the \SiII\ lines, which were already quite strong in
the earlier spectrum. In both this and the previous spectrum, the strength of
the \OI\,7774\,\AA\ line indicates that a significant fraction of the outer
layers was left unburned or only partially burned, as indicated by the absence
of carbon lines \citep{m01}.

\subsection{27 September 2004: $-3$ days}

In the third spectrum, the last before $B$ maximum with an epoch of 16.4 days
after explosion, the trend towards lower temperatures continues, while again
the most apparent evolution is in the strength of the \SII\ lines (Figure 1c).
The IME abundances are not much higher at this depth (9500\,\kms), and reach a
maximum here. The features that characterise the spectrum are the same as at
the previous epoch.  In all the pre-maximum spectra the near-IR continuum is
reproduced reasonably well by the model, which never overestimates the flux by
more than 20\%. This indicates that the opacity, which is mostly due to lines,
is sufficiently large at optical wavelengths that the assumption of an
underlying black body is not grossly incorrect. Line blocking in the UV is
stronger in this spectrum, as the photosphere moves inwards and more metals can
absorb UV photons.

\subsection{2 October 2004: $+2$ days}

The spectra soon after maximum show subtle but significant differences with
respect to the pre-maximum ones. In particular, starting with the spectrum of
2 October 2004, which has an an epoch of 21.4 days after explosion (Figure 1d),
\FeII\  dominates over \FeIII. This is clearly seen in the shape of the feature
near 4800\,\AA, which now shows strong \FeII\ multiplet 48 lines (4923, 5018,
5169\,\AA) and has the characteristic shape of spectroscopically normal but
relatively dim SNe\,Ia. Also, the \OI\ 7774\,\AA\ line became weaker,
indicating that the oxygen content near this rather deep (9000\,\kms)
photosphere is small or zero. The ionization of other ions that determine the
UV also changes, as it does for iron, so that now \CaII, \TiII\ and \CrII,
together with \FeII, determine the UV opacity.  On the other hand, this
spectrum still has a rather blue optical pseudo-continuum.


\begin{figure*}
  \includegraphics[width=144mm]{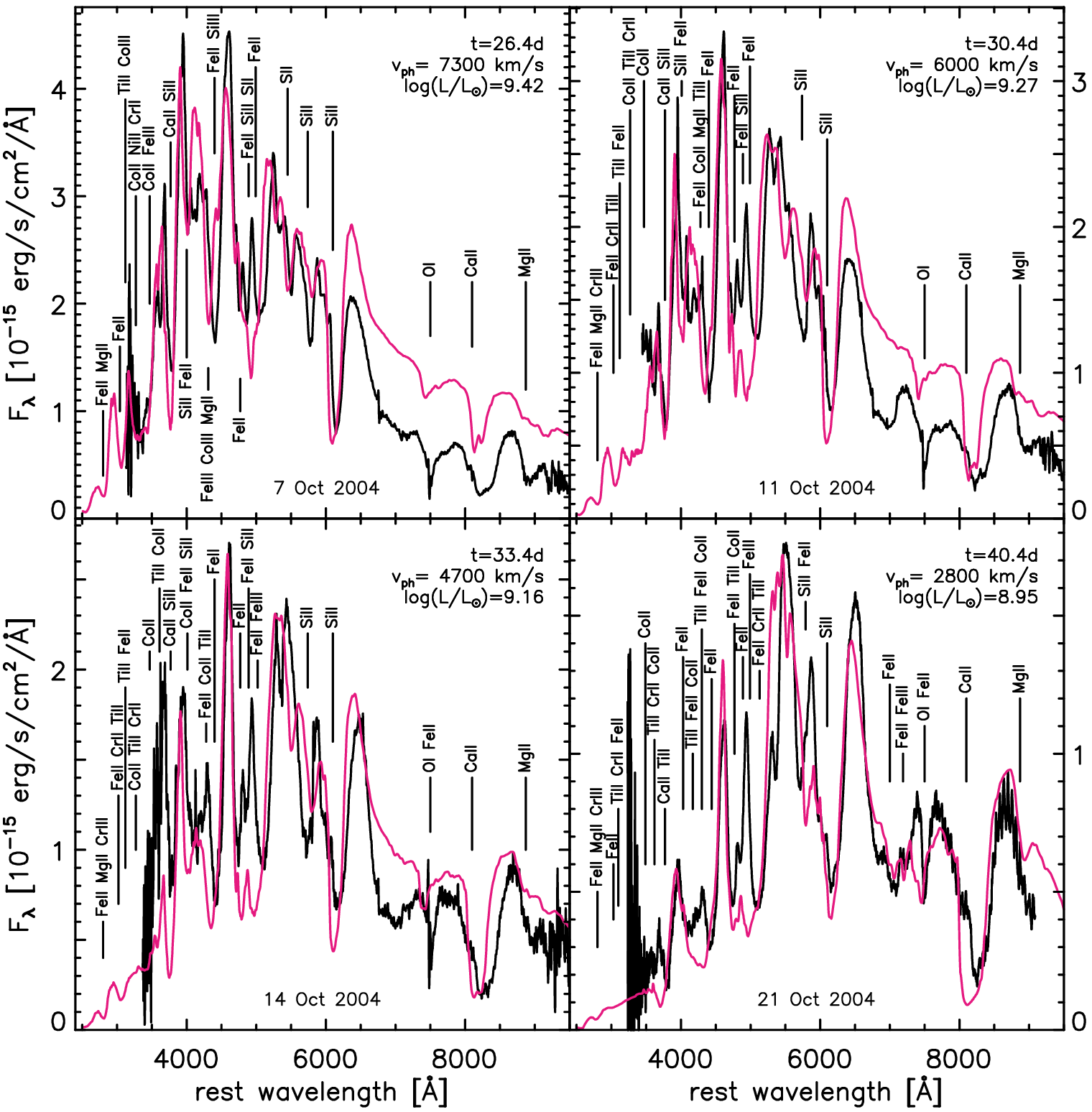}
  \caption{Fits to the spectra of SN\,2004eo in the post-maximum phase.  Basic
  input parameters are shown and and line identifications are marked.  The
  spectra are as follows: {\bf a)} (top left): $t=+7$ days; {\bf b)} (top
  right): $t=+11$ days; {\bf c)} (bottom left): $t=+14$ days; {\bf d)} (bottom
  right): $t=+21$ days.}
\end{figure*}


The model reproduces the observed spectrum at least bluewards of $\sim 6500$\,\AA,
with a near-photospheric composition that is still dominated by IME, but where
\Nifs\ is increasing. The mass enclosed below the photosphere is now
$\sim 0.8 \Msun$, and we are beginning to see the bulk of the NSE material. The
model now overestimates the continuum in the red, indicating that the
assumption of an underlying black body is less appropriate at this epoch.  The
observed pseudo-continuum redwards of the \SiII\,6355\,\AA\ line must be caused
mostly by flux redistribution to the red from line processes that take place in
the blue and UV. Thus, the luminosity used for the model is probably higher
than the real value.  Nevertheless, the strength of the synthetic
\OI\,7774\,\AA\ line is comparable to that of the observed feature. This is
important in order to estimate the oxygen abundance. In general the abundances
are not very different from those of the last pre-maximum spectrum.

\subsection{7 October 2004: $+7$ days}

The next spectrum (Figure 2a) has an epoch of 26.4 days after explosion, and it
continues to show the evolution towards lower temperatures (a larger ratio of
the two \SiII\ lines, the growth of the \SiII-\FeII\ feature near 4800\,\AA)
that started from the earliest phases. The iron absorption is now much
stronger, indicating that we are beginning to see the inner, fully burned zone.
Placing the boundary of the transition between incomplete (IME-dominated) and
complete (NSE-dominated) burning zones somewhere between 7000 and 9000\,\kms\
one derives, using W7, an upper limit to the NSE mass of $\sim 0.7 \Msun$. This
is in agreement with the results of \citet{maz07a}.

\subsection{11 October 2004: $+11$ days}

In this spectrum (Figure 2b), at an epoch of 30.4 days after explosion, the low
temperature begins to affect the pseudo-continuum significantly: the shape of
the spectrum is now much redder. \FeII\ lines have grown in strength. Not only
is the Ni abundance higher, but \Cofs\ is decaying to \Fefs. The fraction of
stable Fe is also larger. This is a feature of SN\,2004eo which is also borne
out by the analysis of the nebular spectra (see Sect. 3). At this and later
epochs, the feature near 5900\,\AA\ is probably contaminated by \NaI\,D.

\subsection{14 October 2004: $+14$ days}

This spectrum (Figure 2c) has a nominal epoch of 33.4 days after explosion, and
is very similar to the one obtained three days earlier. Most features are the
same as in the previous epoch. The \NaI\,D line is growing in strength, as is
typical especially of dim SNe\,Ia. This feature is not reproduced because of
the well-known uncertainty in the sodium ionization \citep[\eg][]{maz97}.
Although the photosphere is now very deep (4700 \kms), the \CaII\ IR triplet
absorption seems to reach even lower velocities, possibly suggesting the need
for a higher calcium abundance at low velocity, or an incorrect estimate of the
ionization there. The composition is dominated by \Nifs\ (60\% by mass) and stable Fe (35\%
by mass), the rest being mostly silicon and sulphur. This agrees with the
nebular spectrum, which samples the region of the ejecta that is near the
photosphere at this epoch.

\subsection{21 October 2004: $+21$ days}

The last of the early-time spectra we modelled has an epoch of 40.4 days after
explosion (Figure 2d). It is very red, and it shows deep and broad absorptions.
The strong emissions are still P-Cygni components caused by the absorption
blends: the epoch is still too early for nebular emission lines to emerge, as
indicated by the good fit to the \CaII\ IR emission component. The broad
features near 4500 and 5000\,\AA\ are caused by \FeII\ lines, although some
\TiII\ contributes to the bluer feature, as it does in very underluminous SNe
\citep{filip92}. The \NaI\,D line is becoming very strong, while other lines
are the same as at earlier epochs.

The synthetic spectrum matches the observed one surprisingly well, considering
the advanced epoch and the very deep photosphere (2800\,\kms), which suggests
that the assumption of a grey photosphere is not unreasonable at this epoch.
The reason for this is probably the fact that since iron dominates the
composition near the photosphere, \FeII, \TiII, and, further to the red, \CoII\
lines effectively block the radiation and form a pseudo-continuum throughout
the optical domain. The composition obtained agrees with that derived from the
late-time spectrum (Sect. 3).

\section{The nebular epoch}

One nebular-epoch spectrum of SN\,2004eo was presented in \citet{pasto07}. It
was obtained 227 days after $B$ maximum, on 16 May 2006. We use this spectrum
to investigate the structure and composition of the inner ejecta. The modelling
results were already included in the compilation of \citet{maz07a}, but the
spectrum has been recalibrated because of scatter in the photometry and the
values are therefore slightly different.

A one-zone version of the nebular code can be used to determine basic
parameters as in \citet{maz98}. The intrinsic width of the emission lines
required to match the broad blends is 7300\,\kms, which is on the low side of
the distribution of line widths in SNe\,Ia \citep{maz98}. Nevertheless there is
good overlap in velocity space between the most advanced photospheric epoch
spectra that we modelled and the nebular spectrum, ensuring complete coverage
of the ejecta and the consistency of the analysis. Material at velocities above
8000\,\kms makes a negligible contribution to the nebular spectrum, indicating
that the densities are too low and that \Nifs\ is not present in any large
amount at those velocities.


\begin{figure}
  \includegraphics{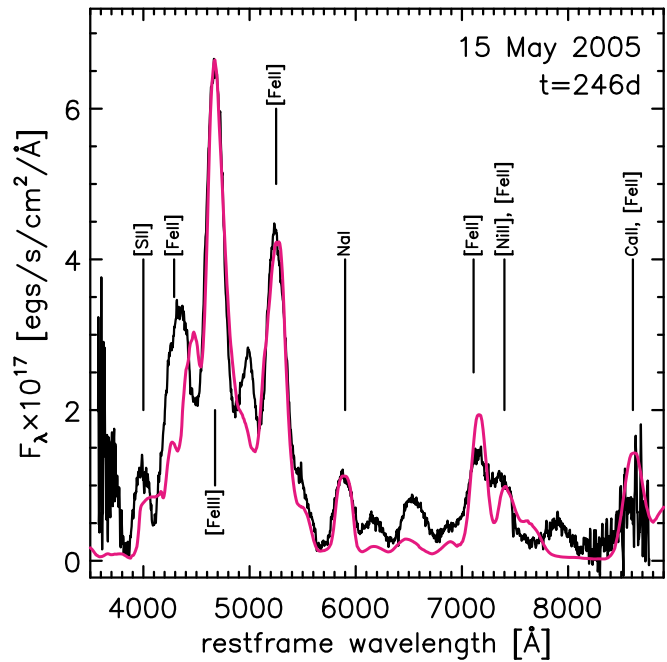}
\caption{The nebular-epoch spectrum of SN\,2004eo and a synthtic match.}
\label{fig:nebspec}
\end{figure}


For the stratified modelling the density stratification of W7 is again adopted,
and the depth-dependent abundances are modified in order to fit the spectrum.

The abundances in the nebula are dominated by NSE material. The best fit
(Figure 3) includes a \Nifs\ mass of $0.38 \Msun$, located mostly between
velocities of 2000 and 8000\,\kms. The abundance of \Nifs\ reaches $\sim 0.65$
by mass between 3000 and 7000\,\kms. The innermost zone is dominated by stable
Fe and Ni. This is necessary in order to avoid the formation of sharp emission
peaks for the strongest [\FeIII] and [\FeII] lines and to achieve an ionization
balance that gives the correct ratio of the [\FeIII]-dominated 4700\,\AA\ line
and the [\FeII]-dominated 5200\,\AA\ line. The mass of stable Fe is $0.27 \Msun$,
so that the total NSE mass is $0.65 \Msun$ (the mass of stable Ni is very
small, in order to avoid a strong [\NiII] emission near 7000\,\AA.


\begin{figure*}
  \includegraphics[width=144mm]{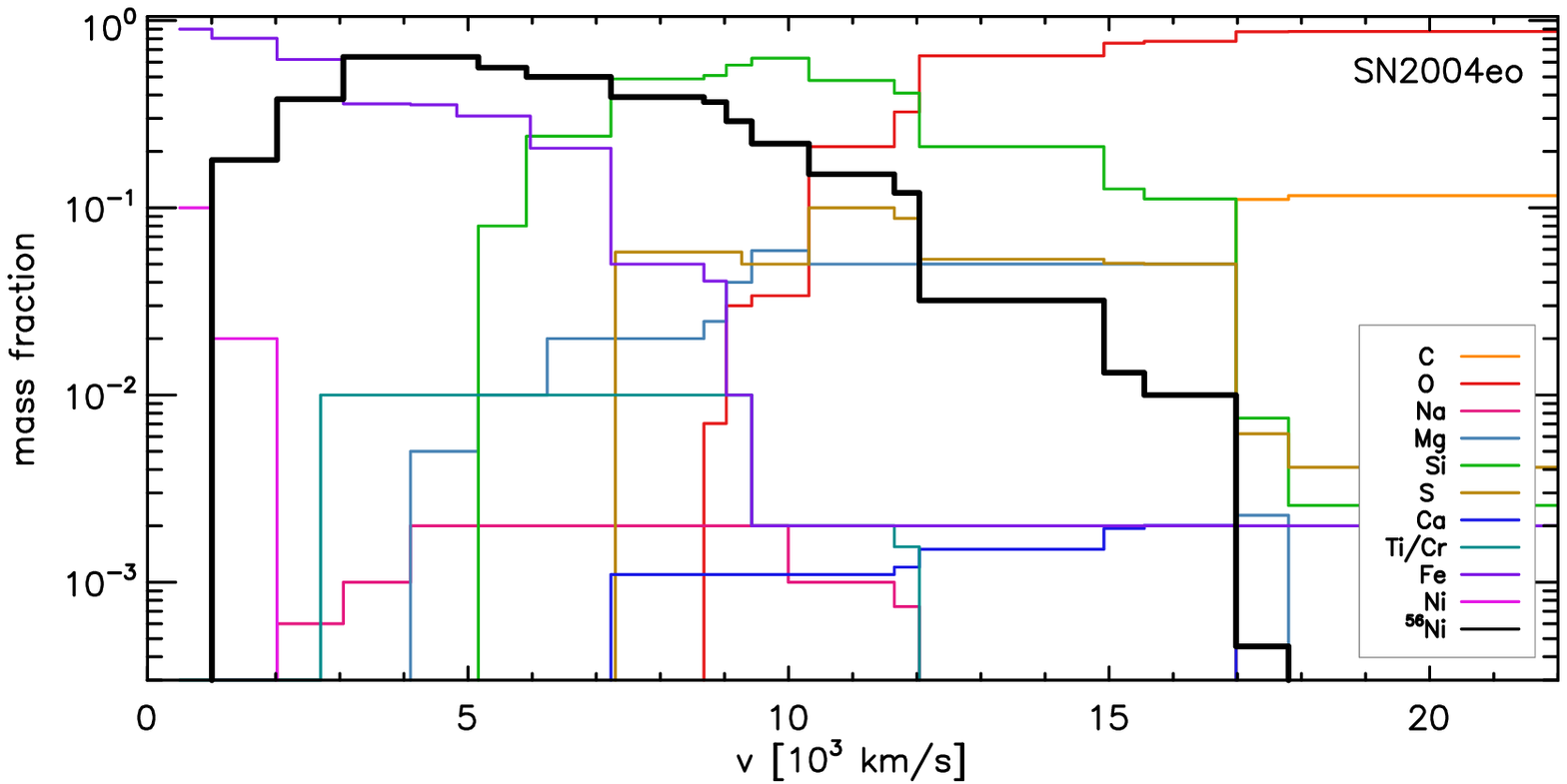}
  \includegraphics[width=144mm]{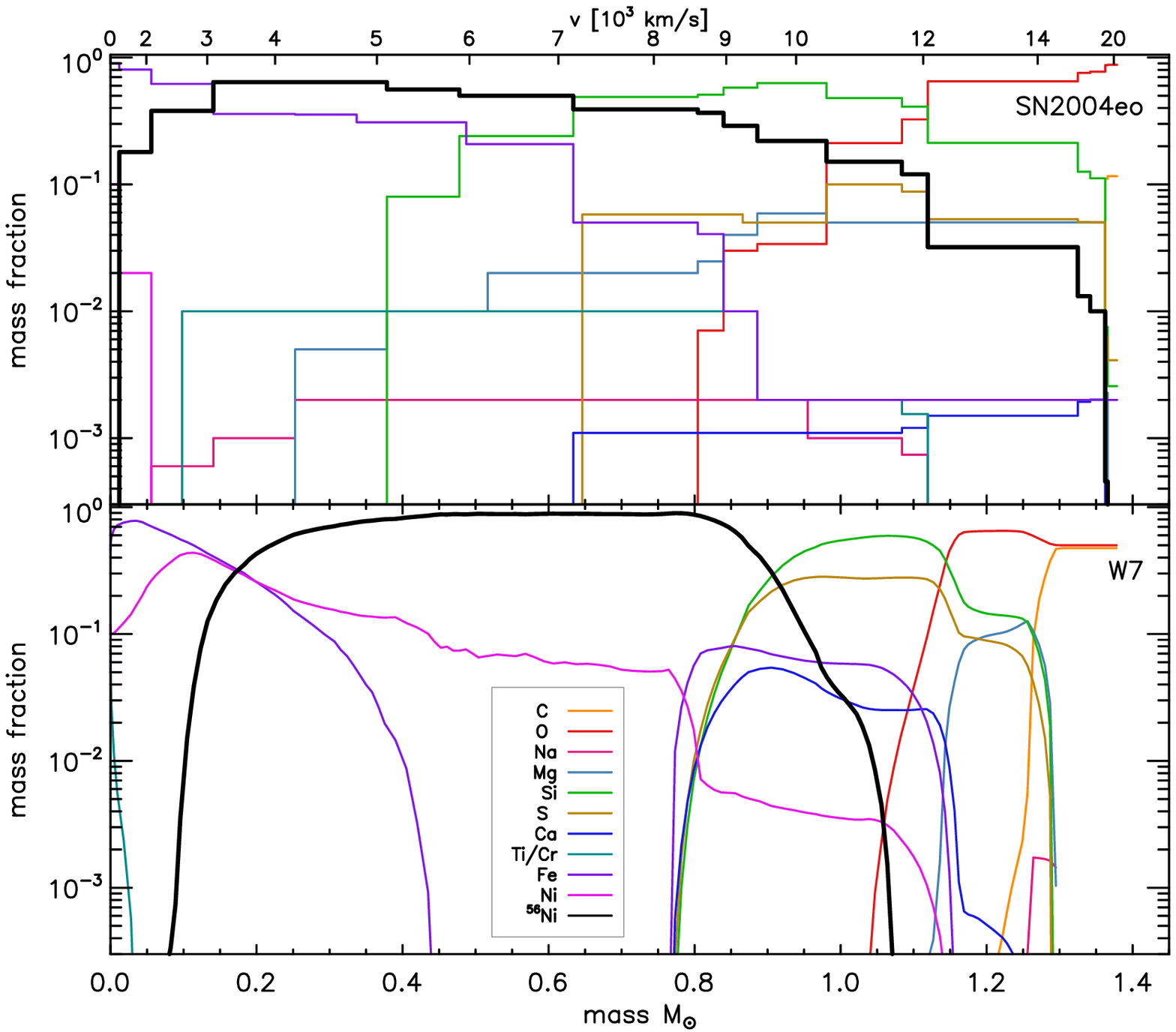}
  \caption{Distribution of the principal elements in the ejecta of SN\,2004eo
  as derived from the spectral fitting.  {\bf a)} [top]: distribution in
  velocity space.  {\bf b)} [middle]: distribution in mass coordinates.  {\bf
  c)} [bottom]: distribution in mass coordinates of the W7 model \citep{nom84}.  }
\end{figure*}


The strongest observed emissions are fitted quite accurately. Many of the
weaker lines that are not reproduced are weak Fe or Co lines, for which the
collision strengths are not well known. The poor reproduction of those features
means that we may be underestimating the \Nifs\ mass by $\sim 5$\%.
Apart from the strong [\FeII] and [\FeIII] lines, \NaI\,D is clearly visible, as
are \CaII\ lines, which are however strongly affected by [\FeII] emission. \MgI]
4700\,\AA\ contributes to the blue wing of the 4700\,\AA\ feature: a Mg mass of
$0.04 \Msun$ is obtained from the fit. Silicon does not have strong lines in the
optical domain, but it has a cooling effect since it produces lines in the IR,
notably at 1.6 microns. A low upper limit for the sulphur abundance can be
established from the line near 4000\,\AA. The oxygen abundance must be very low
at velocities below 8000\kms, or a strong [\OI] 6300\,\AA\ line would form.  The
carbon abundance must also be very low, or [\CI] emission would affect the blue
part of the \CaII\ IR triplet emission. The abundances above 8000\,\kms are the
same as those derived from the early-time modelling. Between 2800 and 7000\,\kms
some difference exists, but this is never more than 10\%.

\section{Abundance stratification}

Connecting the results of early and late-time spectral modelling, we can
reconstruct the abundance distribution as a function of depth in the ejecta.
The results are shown in Figures 4a and 4b. The 11 most abundant species which
contribute to the spectra are displayed. Minor species cannot be treated
accurately owing to the lack of significant spectral features. Figures 4a and
4b show that oxygen dominates the outer ejecta, down to $\sim 12000$\kms. This
is not a particularly deep oxygen zone, despite the overall dimness of
SN\,2004eo, and it is comparable to most other SNe\,Ia \citep{maz07a}. The
absence of carbon is noticeable: no carbon features are present that can be
modelled, and the upper limit as constrained by the early spectra is very low
(a few percent in the outer zones). This is also a feature of most SNe\,Ia,
which may indicate partial burning of carbon to oxygen in the entire progenitor.

IME dominate at velocities between 7000 and 12000\,\kms. Silicon is the most
abundant IME, followed by sulphur and magnesium. The abundances of sodium and
calcium are much smaller. IME are present out to the highest velocities surveyed
by our spectral models, at a level of $\approx 10$\% by mass. They do not extend
significantly below $\sim 6000$\,\kms, indicating again that the innermost part
of the ejecta was completely burned to NSE. Among the NSE isotopes, \Nifs\ is
the most abundant species between 3000 and 7000\,\kms, extending out to
$\sim 12000$\,\kms, while stable NSE isotopes (mostly stable Fe) dominate the
inner 3000\,\kms, as discussed in the previous section.

We can compare this distribution with those of W7 (Figure 4c) and of SN\,2002bo
(Paper I). In SN\,2004eo NSE material does not extend as far out as in
SN\,2002bo, where the bulk reached $\sim 10,000$\,\kms\ but NSE material was
present out to 15,000\,\kms\ at the 10\% level, while W7 is somewhere in between
and shows no mixing-out.  The outer extent of the IME is similar in SNe 2004eo
and 2002bo, confirming the result of \citet{maz07a}. A peculiarity of SN\,2004eo
is the relatively large stable NSE material inner zone.


\begin{figure}
  \includegraphics[width=89mm]{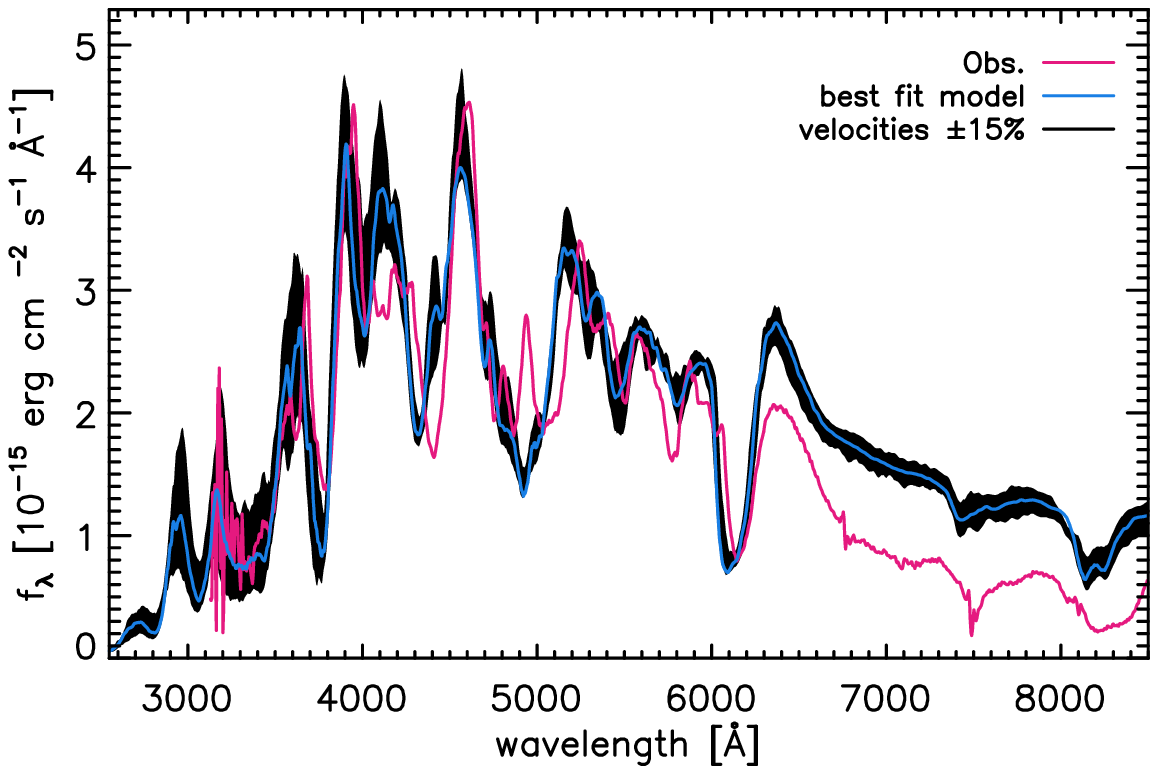}
  \includegraphics[width=89mm]{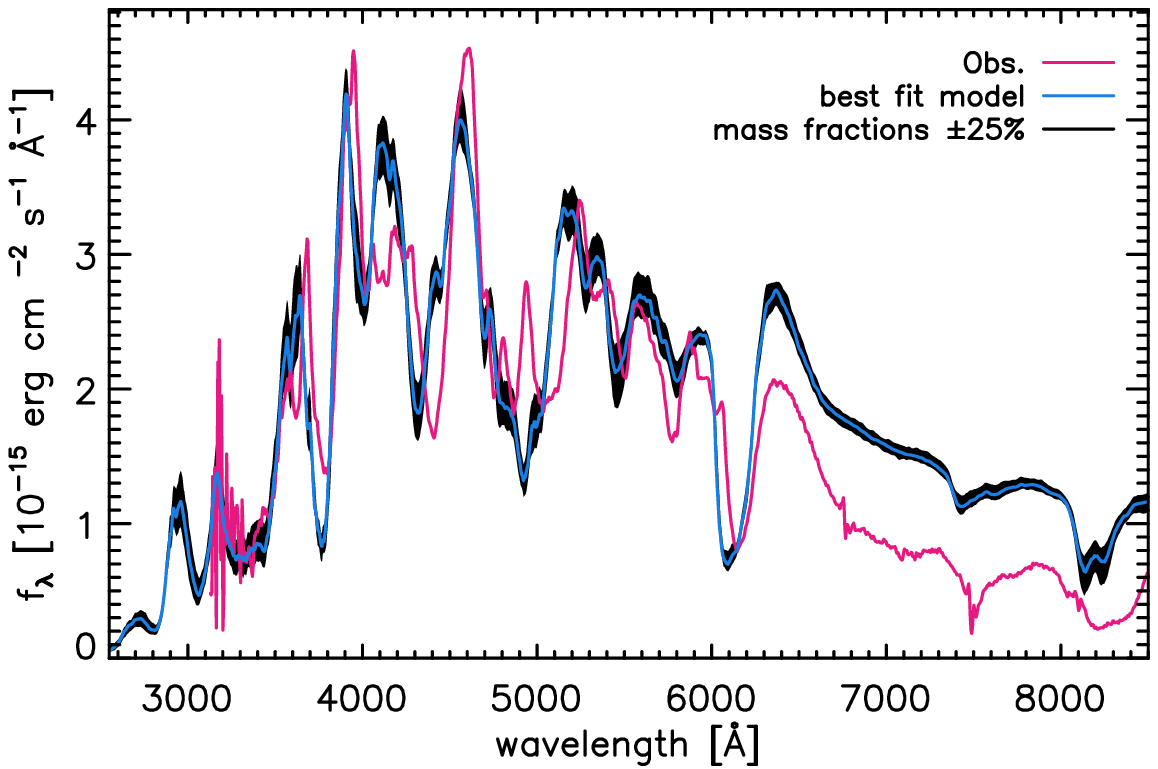}
\caption{{\bf a)} (top) Sensitivity of the results to velocity. The red line
shows the observed $t = 26.4$ day spectrum and the blue line is the best fit.
The black band shows spectra computed changing the velocity boundaries of
the various shells randomly by up to 15\% with respect to the best-fit values.
{\bf  b)} (bottom) Sensitivity to abundances within fixed velocity shells: same
as above, but changing the abundances by up to 25\% within each shell.}
\end{figure}


The abundance distribution we have presented is obtained from our best-fit
models, as discussed in Section 2 above. It is instructive to verify the effect
on the spectral fits of adopting different parameters.  In Figures 5a and 5b we
show the variation induced in the synthetic spectra by randomly varying the
velocity boundaries of the various abundance shells by up to 15\% (Fig. 5a) and
the abundances by up to 25\% (Fig. 5b). A relatively small change in the
position of the shells can affect the result significantly. The sensitivity to
abundance within a fixed shell is smaller, and the uncertainty can be estimated
to be $\sim 25$\% in the photospheric epoch. The examples shown here address the
issue of fitting a particular spectrum. When trying to fit the entire time
series, many of the solutions that may seem acceptable based on a single epoch
do not result in a consistent fit of the spectral evolution. The masses of the
NSE elements are estimated mostly from the nebular model, and have a smaller
uncertainty.

\section{A light curve model}

In Paper I we could show that using the abundances derived from the spectral
analysis to estimate the opacity in the ejecta a more accurate synthetic light
curve could be obtained than any based on theoretical explosion models. In
particular, the mixing out of \Nifs\ is not predicted in most models, which
therefore fail to reproduce the early rise of the light curve. Unfortunately,
the situation is not quite as good for SN\,2004eo, possibly because there are
fewer very early spectra, which makes it difficult to constrain the abundances
above 15000\,\kms with high accuracy. In Figure 5 the bolometric light curve of
SN\,2004eo \citep{pasto07} is compared to a synthetic bolometric light curve
obtained with a Montecarlo code based on that described in \citet{capp97}. In
the calculation, the optical photons produced by the deposition of $\gamma$-rays
and positrons diffuse in the ejecta, where they encounter an optical opacity
which is determined by the abundance of the different isotopic species as follows:

\begin{equation}
	\kappa_{opt} = 0.1 [M({\rm NSE})/\Msun] + 0.01 [M({\rm IME})/\Msun]
\end{equation}

The synthetic light curve based on the abundance distribution derived from
spectrum synthesis (Fig. 5, solid curve) resembles the observed one, with a peak
at $\sim 19$ days, but it lies below the "observed" one by $\sim 0.2$\,mag at
all epochs before $\sim 30$ days, although the tail is well reproduced.

The error introduced by the assumption of a sharp photosphere in the models for
the photospheric-epoch spectra can be assessed by verifying in the synthetic
light curve what fraction of the $\gamma$-ray deposition occurs above the
momentary photosphere.  For example, at maximum light less than 10\% of the
deposition occurs above the photosphere, which is then located at $\sim
9000$\,\kms, above the bulk of \Nifs. This fraction increases to $\sim 50$\% by
day 40, when the photosphere is well within the \Nifs\ zone. However, the
nebular spectrum covers velocities out to $\sim 9000$\,\kms, and so the
uncertainty caused by the use of a sharp photosphere is limited.


\begin{figure}
  \includegraphics[width=89mm]{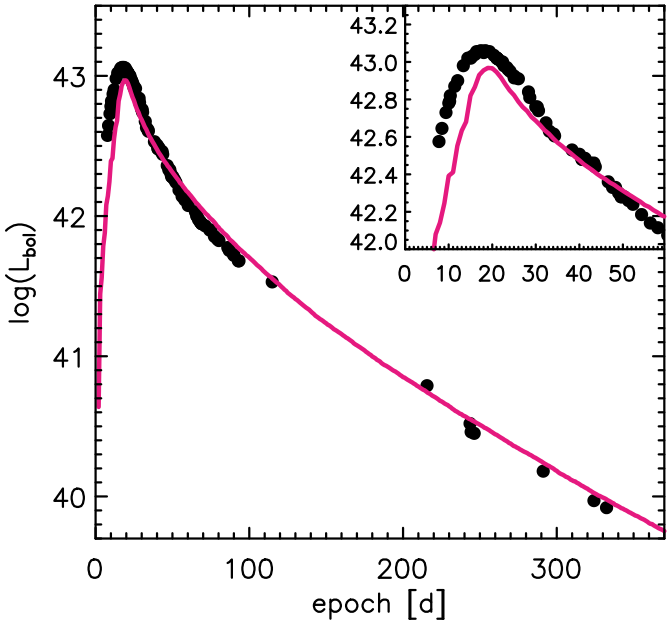}
  \caption{The bolometric light curve of SN\,2004eo \citep{pasto07} compared to
  a synthetic light curve obtained using our Montecarlo light curve model
  \citep{capp97} and the abundances obtained from the spectral fitting discused
  in this paper. Although the synthetic light curve resembles the data, it
  fails to reach the same peak luminosity.  The inset is an enlargement of the
  epoch of maximum. }
\end{figure}


One of the possible reasons for the discrepancy betwen the synthetic and
observed light curves is the estimate of the mass and distribution of \Nifs.

The synthetic light curve can be improved by changing the \Nifs\ mass and
distribution. Some possibilities are shown in Figure 6.  First, if the amount
of \Nifs\ in the outer shells was underestimated because of poor spectral
coverage or uncertainties in the spectral modelling at early times, a different
rising part could be obtained. Adding \Nifs\ at velocities above 10000\,\kms\
improves the rising part up to $\sim$ one week before maximum, but it still
does not match the peak.  In order further to improve the fit, it is necessary
to play with the distribution of \Nifs\ in a way that is not justified by the
spectroscopic analysis.  A model with enhanced \Nifs\ in the region between
7000 and 10000\,\kms\ (Fig. 6, dashed line) results in a better fit of the
peak. This model has $M({\rm ^{56}Ni}) = 0.49 \Msun$, and
$M({\rm NSE}) = 0.77 \Msun$.  However, because of the very large \Nifs\ mass
this model is much too bright on the tail.

One further epicycle is then to remove \Nifs\ at low velocities, below
6000\kms.  This model (Fig. 6, dotted line), with a total \Nifs\ mass of
$0.45 \Msun$ and $M(\rm{NSE}) = 0.76 \Msun$, fits the entire light curve, but
is not physically justified. For example, a nebular spectrum based on this
distribution of \Nifs\ would not at all resemble the observations.


\begin{figure}
  \includegraphics[width=89mm]{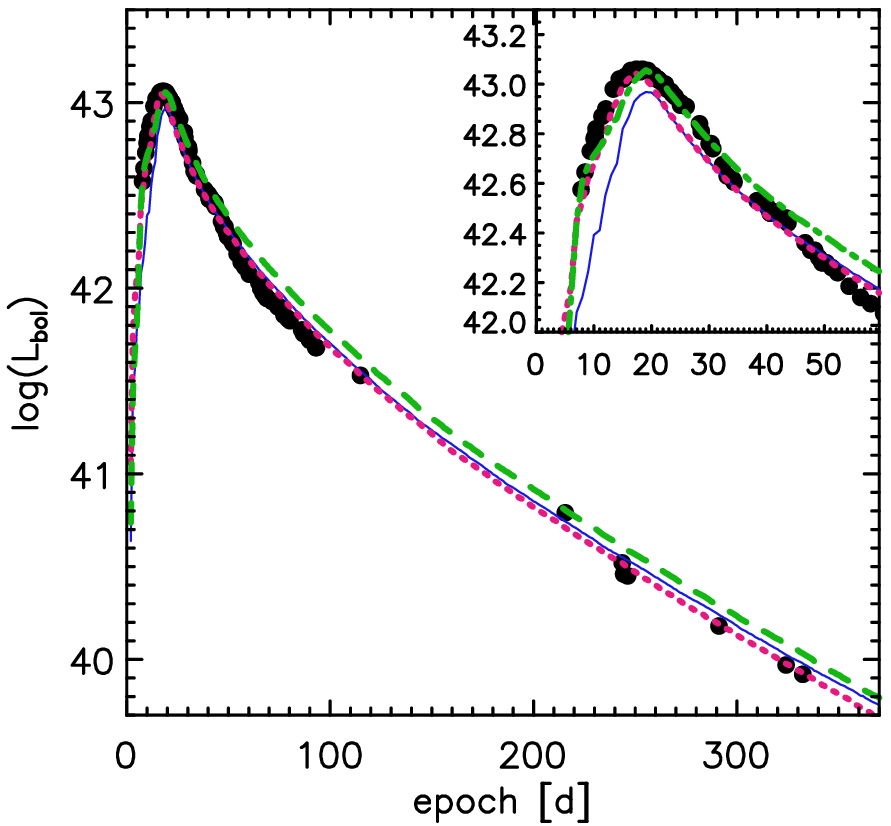}
  \caption{The bolometric light curve of SN\,2004eo \citep{pasto07} compared to
  two synthetic light curves obtained by arbitrarily modifying the mass and
  distribution of \Nifs. In the model shown as a dashed green line \Nifs\ was
  added at velocities between 7000 and 10000\,\kms, for a total M(\Nifs) $= 0.49
  \Msun$. This model fits the peak of the light curve, but is too bright on the
  tail.  In the model shown as a dotted red line the additional \Nifs\ at
  intermdiate velocities was compensated for by removing \Nifs\ at velocities
  below 6000\kms, for a total M(\Nifs) $= 0.45 \Msun$. This model provides a good
  fit to the light curve but it is not physically motivated. The model from the
  abundance stratification [M(\Nifs) $= 0.38 \Msun]$ is also shown for comparison
  (thin blue line). }
\end{figure}


The question is then what is the source of the discrepancy. There are at least 4
possibilities:

1) Calibration of the nebular spectrum. Incorrect calibration would result in
an incorrect estimate of the \Nifs\ content in the inner ejecta, and hence in
an incorrect light curve. This is a possibility since there is scatter in the
photometry obtained near the epoch of the nebular spectrum \citep{pasto07}. It
is however unlikely to result in a different relative distribution of \Nifs\
and stable Fe-group isotopes, as this would directly affect the ionization
balance and hence the line strength ratios.

2) Estimate of the amount of \Nifs\ in the region between 7000 and 10000\kms.
The early-time spectra are relatively insensitive to variations of the \Nifs\
mass in this velocity range: these layers are visible above the photosphere
only after maximum, when most \Nifs\ has decayed to \Cofs\ but only a small
fraction of this has decayed to \Fefs. Fe lines are therefore not greatly
affected, and the effect of replacing nickel with cobalt is small. Nebular
spectra, on the other hand, are also not very sensitive to small variations of
the \Nifs\ content at these velocities because of the low overall density.

3) Estimate of the bolometric correction and its evolution. Evidence for this
may be the fact that even though the spectral fits at maximum require an input
luminosity comparable to or somewhat lower than the corresponding $L_{Bol}$,
the synthetic spectra thus obtained match the optical part of the observed
spectra but are too bright in the near IR. Thus, the real value of $L_{Bol}$ at
peak may actually be smaller than estimated in the bolometric light curve.

4) Uncertain treatment of the opacity. This is not likely to be a problem at
late epochs, when the ejecta are transparent, but it may affect the light curve
calculation near maximum. On the other hand, our synthetic light curve tracks
the observed one quite well, the only problem being that it lies always below
the data. This suggests that if there is a problem with the opacities, it lies
in the $\gamma$-ray opacity, but this is a quantity that is well known and is
unlikely to be grossly in error. The part of the light curve where the optical
opacity is actually very likely to be incorrectly treated is between roughly 50
and 100 days, when the SN is making the transition to the nebular phase. This
however has no effect on the estimate of the mass of \Nifs\ either at the peak
or on the tail of the light curve.

Alternatively, a direct estimate of the mass of \Nifs\ may be attempted using
the prescription of \citet{stritz06}: $L = 2\times10^{43} M({\rm ^{56}Ni})$.
With a bolometric light curve peak of $1.15\times10^{43}$\,erg\,s$^{-1}$, this
yields $M({\rm ^{56}Ni}) = 0.57 \Msun$. This seems a very large value, given
the results of the various light curve tests and the nebular spectrum. If we
include \Nifs\ in the inner part of the ejecta, as suggested by the
spectroscopic modelling and by most explosion models, we can accommodate at
most $0.49 \Msun$ of \Nifs. This yields a good fit to the peak of the light
curve, but it is too bright on the tail and it cannot reproduce the nebular
spectra. Additionally, in all light curve models we have presented, we obtain a
relation $L \sim 2.5\times10^{43} M({\rm ^{56}Ni})$, which again suggests that
$M({\rm ^{56}Ni}) = 0.45 \Msun$. The higher efficiency of conversion of $\gamma$-ray
energy to optical light in our models with respect to the assumptions of
\citet{stritz06} depends essentially on adopting a realistic density profile
and a distribution of \Nifs\ in the ejecta. Given all of the above
uncertainties, a conservative estimate of the \Nifs\ mass of SN\,2004eo is
$M({\rm ^{56}Ni}) = 0.43 \pm 0.05 \Msun$.

\section{Conclusions}

Despite the relatively small number of spectra available for modelling, a
thorough description of the composition of the ejecta of SN\,2004eo was
obtained. This SN had a small \Nifs\ mass for a normal SN\,Ia. The exact value
is made somewhat uncertain by difficulties with deriving the bolometric light
curve as well as by uncertainties in the calibration of the nebular spectrum
and the behaviour of the optical opacity. Still our estimate,
$M({\rm ^{56}Ni}) = 0.43 \pm 0.05 \Msun$, places SN\,2004eo among the least
luminous normal SNe\,Ia, along with the template SN\,1992A.

A possible caveat concerns the total mass of the ejecta and the energy of the
explosion. The assumption of the Chandrasekhar mass and of the W7 model yields
a rather sensible abundance pattern and reasonably good synthetic spectra,
suggesting that any deviation from that model should not be very large. Had the
mass been significantly smaller, in our W7-based models we would have had to
``hide'' mass in elements that make a small contribution to the spectra: most
likely oxygen in the photospheric phase and silicon in the nebular phase. This
was however not required, and most elements included in the mixture do
contribute significantly to the spectra at the appropriate epochs, via
absorption lines at early times or emission lines in the nebular phase. As for
the kinetic energy, applying the formula of \citet{woosetal07},
$E_K = [1.56 M({\rm ^{56}Ni}) + 1.74 M({\rm stable NSE}) + 1.24 M({\rm IME}) - 0.46] 10^{51}$
\,erg, we obtain $E_K = 1.1 \pm 0.1\times10^{51}$\,erg. This is somewhat less
than the original W7 energy ($1.3\times10^{51}$\,erg), but not so much less:
the smaller contribution of burning to \Nifs\ is in fact compensated by the
larger production of IME, which contribute to the kinetic energy only $\sim
20$\% less per unit mass synthesised, and by the large production of stable NSE
isotopes, which actually contribute more to the kinetic energy than burning to
\Nifs. The main uncertainty regarding the kinetic energy comes from the outer
extent of burning to NSE and the degree to which the outer carbon and oxygen
were burned to IME. The zone between 9000 and 11000 \kms still contains some
oxygen.  A signature of the lack of incomplete burning products at high
velocities may be seen in the lack of HVF's, at least at the epochs sampled by
our spectra.

In conclusion, within the uncertainties SN\,2004eo was a Chandrasekhar-mass
explosion of a CO white dwarf that produced a smaller amount of \Nifs\ than
average ($\sim 0.43 \Msun$). The amount of stable Fe-group material
($\sim 0.25 \Msun$), is somewhat larger than average \citep{maz07a}. The outer
extent of IME ($\sim 1.1 \Msun$) is similar to most SNe\,Ia [the exception
being those defined as High Velocity Gradient by \citet{ben05}], so that the
IME production was larger than average ($\sim 0.4 \Msun$). Since there is no
indication of unburned carbon, most of the remaining $\sim 0.4 \Msun$ of
material should be unburned oxygen.

Such a low luminosity explosion may be the result of a strong initial
deflagration phase, during which NSE material is produced and the star
pre-expands, so that if and when a delayed detonation occurs this is not very
effective, burning material at low density mostly to IME only \citep{maz07a}.
Uncertainties arise mostly from the spectral calibration, the bolometric
correction and the lack of extremely early data, calling for more complete
coverage of SNe\,Ia so that the parameters of the explosions can be derived
with even greater precision. Nevertheless, SN\,2004eo is an important piece in
the description of the variety of SNe\,Ia and in the effort to understand their
origin.

\vspace{3mm}
\noindent
This research was supported by the European Union's Human Potential Programme
under contract HPRN-CT-2002-00303 ``The Physics of Type Ia Supernovae" and by
the National Science Foundation under Grant No. PHY05-51164. D.S. was supported
by the Transregional Research Center TRR33 "The Dark Universe" of the DFG.





\begin{thebibliography}{}

\bibitem[Arnett(1982)]{arn82} Arnett, D. 1982, \apj, 253, 785


\bibitem[Axelrod(1980)]{axe80} Axelrod, T.~S.\ 1980, Ph.D.~Thesis, Univ. of
    California, Santa Cruz

\bibitem[Benetti et al.(2004)]{ben04} Benetti, S., et al.\ 2004, \mnras, 348,
    261

\bibitem[Benetti et al.(2005)]{ben05} Benetti, S., et al.\ 2005, \apj, 623,
    1011


\bibitem[Cappellaro et al.(1997)]{capp97} Cappellaro, E., Mazzali, P.~A.,
    Benetti, S., Danziger, I.~J., Turatto, M., della Valle, M.,
    \& Patat, F.\ 1997, \aap, 328, 203

\bibitem[Contardo, Leibundgut, \& Vacca(2000)]{cont00} Contardo, G.,
    Leibundgut, B., Vacca, W.D. 2000, \aap, 359, 876



\bibitem[Filippenko et al.(1992)]{filip92} Filippenko, A.\,V., et al.\ 1992,
    \aj, 104, 1543

\bibitem[Garnavich et al.(2004)]{garn04} Garnavich, P.\,M., et al.\ 2004,
    \apj, 613, 1120

\bibitem[Hachinger et al.(2008)]{hach08} Hachinger, S., et al.\ 2008, \mnras,
    submitted




\bibitem[Hillebrandt \& Niemeyer(2000)]{HillNie00} Hillebrandt, W.,
    \& Niemeyer, J.\,C.\ 2000, \araa, 38, 191

\bibitem[Hoeflich et al.(1993)]{hmk93} Hoeflich, P., Mueller, E.,
    \& Khokhlov, A.\ 1993, \aap, 268, 570



\bibitem[Karp et al.(1977)]{karp77} Karp, A.~H., Lasher, G., Chan, K.~L.,
    \& Salpeter, E.~E.\ 1977, \apj, 214, 161

\bibitem[Khokhlov(1991)]{khok91} Khokhlov, A.~M.\ 1991, \aap, 245, 114


\bibitem[Kuchner et al.(1994)]{kuch94} Kuchner, M.~J., Kirshner, R.~P.,
    Pinto, P.~A., \& Leibundgut, B.\ 1994, \apjl, 426, L89

\bibitem[Iwamoto et al.(1999)]{iwa99} Iwamoto, K., Brachwitz, F., Nomoto, K.,
    Kishimoto, N., Umeda, H., Hix, W.~R., \& Thielemann, F.-K.\ 1999,
    \apjs, 125, 439


\bibitem[Lucy(1999)]{l99} Lucy,~L.B. 1999, \aap, 345, 211

\bibitem[Mazzali(2000)]{m00} Mazzali,~P.A. 2000, \aap, 363, 705

\bibitem[Mazzali(2001)]{m01} Mazzali, P.~A.\ 2001, \mnras, 321, 341

\bibitem[Mazzali et al.(1998)]{maz98} Mazzali, P.A., Cappellaro, E.,
    Danziger, I.J., Turatto, M, Benetti, S. 1998, \apj, 499, L49

\bibitem[Mazzali et al.(1997)]{maz97} Mazzali, P.\,A., Chugai, N., Turatto, M.,
    Lucy, L.\,B., Danziger, I.\,J., Cappellaro, E., della Valle, M.,
    \& Benetti, S.\ 1997, \mnras, 284, 151

\bibitem[Mazzali \& Lucy(1993)]{m&l93} Mazzali,~P.A., Lucy,~L.B. 1993,
    \aap, 279, 447

\bibitem[Mazzali et al.(1993)]{maz93} Mazzali, P.~A., Lucy, L.~B.,
    Danziger, I.~J., Gouiffes, C., Cappellaro, E., \& Turatto, M.\ 1993,
    \aap, 269, 423

\bibitem[Mazzali et al.(2005)]{maz05} Mazzali, P.~A., et al.\ 2005,
    \apjl, 623, L37

\bibitem[Mazzali et al.(2001)]{maz01} Mazzali, P.~A., Nomoto, K.,
    Cappellaro, E., Nakamura, T., Umeda, H., Iwamoto, K. 2001, \apj, 547, 988

\bibitem[Mazzali \& Podsiadlowski(2006)]{m&p06} Mazzali, P.~A., \&
    Podsiadlowski, Ph.\ 2006, \mnras, 369, L19

\bibitem[Mazzali et al.(2007a)]{maz07a} Mazzali, P.~A., R{\"o}pke, F.~K.,
    Benetti, S., \& Hillebrandt, W.\ 2007, Science, 315, 825

\bibitem[Mazzali et al.(2007b)]{maz07b} Mazzali, P.~A., et al.\ 2007, \apj,
    670, 592

\bibitem[Nomoto et al.(1984)]{nom84} Nomoto, K., Thielemann, F.-K.,
    Yokoi, K. 1984, \apj, 286, 644

\bibitem[Nugent et al.(1995)]{nug95} Nugent, P., Phillips, M., Baron, E.,
    Branch, D., Hauschild, P.H. 1995, \apj, 455, L147

\bibitem[Pastorello et al.(2007)]{pasto07} Pastorello, A., et al.\ 2007,
    \mnras, 377, 1531

\bibitem[Pauldrach et al.(1996)]{paul96} Pauldrach, A.W.A., Duschinger, M.,
    Mazzali, P.A., et al.\ 1996, \aap, 312, 525

\bibitem[Perlmutter et al.(1997)]{perl99} Perlmutter, S., et al.\ 1997, \apj,
    483, 565

\bibitem[Phillips(1993)]{phil93} Phillips, M.M. 1993, \apj, 413, L105

\bibitem[Phillips et al.(1999)]{phil99} Phillips, M.M., Lira, P.,
    Suntzeff, N.B., Schommer, R.A., Hamuy, M., Maza, J. 1999, \aj, 118, 1766

\bibitem[Pinto \& Eastman(2000)]{p&e00} Pinto, P.A., \& Eastman, R.G. 2000,
    \apj, 530, 757

\bibitem[Riess et al.(1996)]{riesspresskirsh96} Riess, A.~G., Press, W.~H.,
    \& Kirshner, R.~P.\ 1996, \apj, 473, 88

\bibitem[Riess et al.(1998)]{riess98} Riess, A.~G., et al.\ 1998,
    \aj, 116, 1009

\bibitem[Riess et al.(1999)]{riess99} Riess, A.~G., et al.\ 1999,
    \aj, 118, 2675


\bibitem[R{\"o}pke(2007)]{roep07} R{\"o}pke, F.\,K.\ 2007, \apj, 668, 1103

\bibitem[R{\"o}pke et al.(2007)]{roepke07} R{\"o}pke, F.\,K., Hillebrandt, W.,
    Schmidt, W., Niemeyer, J.\,C., Blinnikov, S.\,I., \& Mazzali, P.\,A.\ 2007,
    \apj, 668, 1132

\bibitem[R{\"o}pke \& Niemeyer(2007)]{roepniem07} R{\"o}pke, F.\,K. \&
    Niemeyer, J.\,C.\ 2007, \aap, 464, 683

\bibitem[Ruiz-Lapuente \& Lucy(1992)]{rll92} Ruiz-Lapuente, P.,
    \& Lucy, L.~B.\ 1992, \apj, 400, 127

\bibitem[Salvo et al.(2001)]{salvo01} Salvo, M.\,E., Cappellaro, E.,
    Mazzali, P.\,A., Benetti, S., Danziger, I.\,J., Patat, F., \& Turatto, M.\
    2001, \mnras, 321, 254

\bibitem[Stehle et al.(2005)]{steh05} Stehle, M., Mazzali, P.\,A.,
    Benetti, S., \& Hillebrandt, W.\ 2005, \mnras, 360, 1231 (Paper I)

\bibitem[Stritzinger et al.(2006)]{stritz06} Stritzinger, M.,
    Leibundgut, B., Walch, S., \& Contardo, G.\ 2006, \aap, 450, 241

\bibitem[Tanaka et al.(2008)]{tan08} Tanaka, M., et al. 2008, \apj, in press

\bibitem[Timmes, Brown, \& Truran(2003)]{timbt03} Timmes, R., Brown, E.F.,
    Truran, J.W. 2003, \apj, 590, L83






\bibitem[Woosley(2007)]{woos07} Woosley, S.~E.\ 2007, \apj, 668, 1109

\bibitem[Woosley et al.(2007)]{woosetal07} Woosley, S.\,E., Kasen, D.,
    Blinnikov, S., \& Sorokina, E.\ 2007, \apj, 662, 487

\end{thebibliography}
\end{document}